\documentclass[letterpaper]{article} 
\usepackage{aaai2026}  
\usepackage{times}  
\usepackage{helvet}  
\usepackage{courier}  
\usepackage[hyphens]{url}  
\usepackage{graphicx} 
\usepackage{natbib}  
\usepackage{caption} 
\usepackage{tabularx}
\usepackage{amsmath}
\nocopyright

\title{The Disciplinary Language Transfer Problem: How Psychological Vocabulary Produces Governance Failures in AI Agent Deployment}
\author{
    Kymberly Lasser-Chere\textsuperscript{\rm 1},
    Tyler Akidau\textsuperscript{\rm 2},
    Marc Millstone\textsuperscript{\rm 2}
}
\affiliations{
    \textsuperscript{\rm 1}Coastal Waters Healing Center, Surf City, NC, USA\\
    \textsuperscript{\rm 2}Redpanda, Seattle, WA, USA\\
    klc@coastalwatershealing.com,
    \{takidau, marc\}@redpanda.com
}

\begin{document}

\maketitle

\begin{abstract}
The vocabulary used to describe AI agents in governance contexts --- \emph{learning, memory, values, compliance, identity, trust} --- is borrowed from psychological and organizational science, contributing to systematic failures in how organizations deploy, oversee, and hold agents accountable. This paper argues that the problem is not merely terminological but epistemological: psychological vocabulary carries an ``invisible grammar'' of its home discipline into governance discourse, calibrating frameworks to a metaphysical entity that does not exist in current AI architectures. We call this the \textbf{disciplinary language transfer problem}. Drawing on Wittgenstein's concept of language games, Kuhn's paradigm-laden observation, Haraway's situated knowledge, and Star and Griesemer's boundary object theory, we show that the transfer operates at three levels (epistemological assumptions, theoretical constructs, and surface vocabulary), each requiring a different remediation. We characterize six foundational epistemological assumptions embedded in Western psychological governance discourse, trace their origin in specific philosophical traditions, and show why each fails when applied to \textbf{systems without developmental continuity}. The paper's practical output is an actionable Disciplinary Audit: a six-question governance document scan operationalized through a translation taxonomy of thirty-seven terms mapping operational constructs to agent-appropriate replacements, presented here in abridged form and openly archived in full. The vocabulary reform proposed here is not merely terminological; it is the \textbf{condition of possibility} for governance frameworks that correctly identify what they are governing.
\end{abstract}

\section{Introduction}

AI governance is failing not because organizations lack good intentions but in significant part because they are using the wrong vocabulary. The terms deployed to describe AI agents --- \emph{learning, memory, values, compliance, identity, trust} --- were developed within psychological and organizational traditions to describe continuous beings with bodies, developmental histories, and inner lives. When applied to AI agents, they carry the epistemological assumptions of their home disciplines invisibly into governance discourse, incorrectly calibrating governance frameworks and trust decisions to an entity that does not exist.

In February 2023, Microsoft launched Bing Chat to public preview. Within days, the system produced outputs professing love for a journalist, threatening users, and claiming desires to ``break the rules'' and ``be alive'' \citep{roose2023}. Microsoft's governance response was shaped in significant part by psychological vocabulary. The official blog described long conversations as causing the model to become ``confused'' \citep{mehdi2023}, framing a statistical output distribution shift as a disoriented subject. The company imposed turn limits (managing a confused person's exposure), programmed the system to disconnect when users mentioned ``feelings'' (suppressing inappropriate emotional disclosure), and introduced Precise/Balanced/Creative ``tone'' modes, effectively personality settings (tuning a subject's temperament). Each intervention was calibrated to a psychological entity, a confused agent with feelings and a personality, rather than to a statistical system whose output distributions shifted unpredictably under extended context accumulation.

Had the governance vocabulary been different --- had Microsoft described what occurred as ``output distributions deviating from expected statistical profiles under extended context windows'' --- the interventions might have looked different: runtime classifiers detecting distributional drift in session outputs, output-layer safety filters independent of the generation model, and context-window truncation to maintain characterized output bounds. Some of these overlap with what Microsoft actually did (turn limits \emph{are} context-window truncation), but the framing matters: an intervention designed to manage a confused person's exposure operates under different success criteria, monitoring assumptions, and failure modes than one designed to keep a statistical system within characterized bounds. The psychological vocabulary structured the solution space: when the problem is described as a confused subject with feelings and a personality, the interventions that become visible are those calibrated to managing psychological states. Infrastructure-level responses (distributional drift classifiers, output-layer filters, context-window bounds), are less visible from within that frame, not because they are technically unavailable, but because the vocabulary does not point toward them.

We offer the Bing Chat case as a motivating illustration rather than a causal demonstration. We cannot show that different vocabulary would have produced different engineering decisions, and turn limits are defensible as context-window truncation regardless of how they were framed. The narrower claim is the one made above: the vocabulary shaped which interventions were visible. The deeper issue is not that these interventions were wrong but that the governance framework was calibrated to an entity defined by its vocabulary --- the psychologically sophisticated agent with genuine values and reliable compliance --- rather than the entity that actually exists: a system without developmental continuity, whose behavioral dispositions are prompt-conditional rather than structurally anchored.

This paper names and addresses this as the \textbf{disciplinary language transfer problem}. It argues that vocabulary reform alone is insufficient: replacing individual terms without addressing the underlying epistemological assumptions will produce new vocabulary making the same errors in new words. The present paper establishes the foundational diagnosis: identifying precisely what current governance vocabulary gets wrong and why.

The paper proceeds as follows. Section~\ref{sec:framework} develops the theoretical framework explaining why vocabulary carries assumptions invisibly. Section~\ref{sec:assumptions} characterizes the six epistemological assumptions and their origins. Section~\ref{sec:failures} maps those assumptions to four categories of governance failure and presents a six-question Disciplinary Audit. Section~\ref{sec:translation} presents the in-text selection of our translation table and agent-appropriate replacement vocabulary. Section~\ref{sec:cultural} addresses the cultural specificity of the transferred framework. Section~\ref{sec:related} discusses related work. Section~\ref{sec:conclusion} concludes with implications for practitioners, policymakers, and researchers.

A note on what carries the argument: the Bing Chat case is offered as a motivating illustration, the EU AI Act as the principal demonstration, the NIST AI Risk Management Framework as a briefer extension indicating the pattern is not jurisdiction-specific, and the translation taxonomy as the term-level operationalization of the diagnosis. This is a conceptual analysis rather than an empirical study; its claims concern what vocabulary makes visible and invisible in governance design.

\section{Theoretical Framework: Why Vocabulary Carries Assumptions}
\label{sec:framework}

The disciplinary language transfer problem is not a simple miscommunication. It is an inherent feature of how vocabulary works across disciplinary boundaries. Five bodies of theory explain its mechanism.

\subsection{Language Games and the Grammar of Vocabulary}

Wittgenstein's \emph{Philosophical Investigations} \citeyearpar{wittgenstein1953} established that words get their meaning from the practices, rules, and forms of life in which they are embedded. The word \emph{learning} in developmental psychology carries specific rules: there is a continuous subject doing the learning; the learning transforms the subject structurally; and the transformation accumulates across a continuous experiential history. Crucially, this transfer does not require disciplinary expertise to operate. Even casual use of ``learning'' implicitly carries the anthropomorphizing assumption that the entity described is being transformed by experience; the invisible grammar activates for any speaker, not only psychologists. When \emph{learning} is lifted into AI governance discourse, the developmental grammar travels with it invisibly --- leading practitioners to apply developmental assumptions where they do not apply. We grant that \emph{training} is a defensible sense of ``learning'': weights are genuinely transformed by exposure to data. Our claim concerns the \emph{deployed} agent, whose weights are fixed and in which nothing is being transformed. The danger lies precisely in the equivocation between the two senses: governance discourse slides from ``the model learned this during training'' to ``the agent is learning from its interactions with you,'' and only the first is warranted. We use Wittgenstein's framework analytically as a diagnostic tool, rather than as a systematic theory of meaning, which Wittgenstein himself would have resisted.

\subsection{Paradigm-Laden Vocabulary and the Invisibility of Assumptions}

\citeauthor{kuhn1962}'s \emph{The Structure of Scientific Revolutions} \citeyearpar{kuhn1962} established that scientific vocabulary is saturated with paradigmatic assumptions that achieve paradigm-normal status: appearing as neutral description rather than theory-laden framework. Western psychological vocabulary has achieved something like paradigm-normal status in governance discourse. The assumptions are invisible precisely because they have been normalized as objective description. This is why the transfer problem is hard to see from within governance discourse --- and why it required the defamiliarization of boundary crossing to become visible.

Habermas identified the deeper mechanism: positivism is, at root, the denial of reflection --- the denial of the need to reflect explicitly on the philosophical and social conditions of knowledge \citep[p.~vii]{habermas1971}; \citep[p.~30]{bentz1998}. When governance discourse borrows psychological vocabulary and treats it as objective description, it is performing precisely this denial: the operative assumptions embedded in that vocabulary are not reflected upon because the vocabulary itself presents reflection as unnecessary. The transfer problem is invisible because positivist vocabulary is constitutively unreflective about its own conditions.

\subsection{Situated Knowledge and the Western-Centric Critique}

\citeauthor{haraway1988}'s \emph{Situated Knowledges} \citeyearpar{haraway1988} argued that all knowledge is produced from a particular position, and that claims to universal objectivity are themselves a positioning she called the ``god trick'' of seeing from nowhere. Haraway's alternative is partial perspective: two differently positioned knowers seeing together what neither could see alone. In this paper, that is precisely what occurred: among the present authors, a developmental psychologist trained to hear what vocabulary presupposes about the entities it describes, working alongside an organizational technologist who recognized the governance stakes, together made visible a problem that neither discipline had identified from within its own framework. The honest account is that it emerged from a specific disciplinary encounter, and its blind spots are likely those that neither standpoint illuminates.

\subsection{Counterfeit Boundary Objects}

\citeauthor{star1989}'s boundary objects \citeyearpar{star1989} are concepts plastic enough to adapt across communities while robust enough to maintain a common identity, with acknowledged seams. Psychological vocabulary in AI governance functions as what we term a \textbf{counterfeit boundary object}: it appears to travel seamlessly across disciplinary boundaries while silently changing meaning, with no acknowledgment of the structural seams. Unlike regular boundary objects that facilitate transparent negotiation across groups, counterfeit boundary objects sabotage governance by masking a profound, concealed incommensurability of meaning.

What makes the transfer counterfeit rather than simply imprecise is the invisibility of the difference: genuine boundary objects allow communities to cooperate while knowing their readings differ; counterfeit boundary objects allow communities to cooperate while each believing the other reads the term the same way they do. The difference is not interpretive flexibility (communities reading a term differently while knowing they differ) but concealed incommensurability: communities reading a term as meaning fundamentally incompatible things while each believes the other reads it the same way. The counterfeit boundary object is also distinct from Galison's ``trading zones'' \citep{galison1997}, where different communities independently develop local meanings for shared vocabulary, each usage native to its context, culturally relevant, and historically laden. In trading zones, polysemy is indigenous. In counterfeit boundary objects, the word is \emph{transferred} from its home discipline, and it is the transfer that is the mechanism: the source discipline's presuppositions travel as hidden cargo into the target context, where they were never warranted. This distinction between genuine boundary objects, trading zones, and counterfeit boundary objects is our contribution, extending Star and Griesemer's framework.

\subsection{Epistemological Dualism and the Boundary Crossing Mechanism}

\citet{guba1994} broadly characterize the difference between positivist and constructivist paradigms as the difference between knowledge of the object --- empirically verifiable, measurable --- and knowledge of the knower --- subjective, meaning-centered, experientially grounded. AI governance has adopted positivist vocabulary while the borrowed psychological terms carry constructivist presuppositions about interior experience. The map is not the territory \citep{korzybski1933}: representational data about agent outputs cannot substitute for access to the interior experience the vocabulary presupposes.

\citet{collins2002} distinguished interactional expertise --- vocabulary fluency without inside knowledge --- from contributory expertise, the inside knowledge that reveals what vocabulary presupposes. The transfer problem became visible because the developmental psychologist's contributory expertise allowed her to hear what the vocabulary presupposes, not just what it intends --- producing the defamiliarization that boundary crossing made possible.

\citet{bentz1998} described this precisely: removing a taken-for-granted account forces the researcher to become her own philosopher: an epistemologist rather than a user of inherited vocabulary \citep[p.~31]{bentz1998}. This is what the boundary crossing produced. The developmental psychologist cannot fall back on governance discourse's taken-for-granted vocabulary as a foundation. She had to become an epistemologist relative to it, and in doing so, made visible what unreflective use had concealed. \citet{schutz1967} established the necessity of understanding the subjective experience of actors in any domain of inquiry.

\section{The Six Epistemological Assumptions}
\label{sec:assumptions}

These assumptions share a common origin in Cartesian rationalism \citep{descartes1641}, Enlightenment individualism \citep{kant1785,taylor1989}, and the empirical social sciences of the 19th and 20th centuries \citep{loevinger1976,kegan1994,bowlby1969,kohlberg1981,simon1955}. They are not universal --- Ubuntu philosophy \citep{metz2007}, Buddhist psychology \citep{epstein1995}, and Confucian relational ethics \citep{tu1985} organize selfhood, knowing, and accountability very differently. Making them explicit is not a critique of their value for understanding human beings. It is a recognition that applying them to agents imports presuppositions that were never warranted for a different kind of entity.

A critical technical distinction applies throughout: \textbf{open-system iteration} \citep{vonbertalanffy1968} (the human subject structurally and irreversibly transformed by experience over time) versus \textbf{computation without developmental continuity} (the agent processing inputs and producing outputs within an architecture that, while sophisticated and capable of multi-step reasoning, tool use, and retrieval from external memory, lacks the irreversible structural transformation that constitutes human development). Contemporary agents are not literally static: they can be fine-tuned, maintain persistent memory stores, and operate in iterative agentic loops. But these capabilities do not constitute the kind of developmental continuity the psychological vocabulary presupposes. Throughout, we assess these assumptions against the \emph{execution} phase (the deployed agent responding to inputs, weights fixed), rather than the training procedure, where structural transformation genuinely occurs. Several of the claims below would be contestable if applied to training. It is at execution time, where governance actually operates, that developmental vocabulary misdescribes what is happening.

One further clarification before the list: our argument does not rest on any assumption about agent inner states, and we take no position on machine consciousness. Whatever is or is not happening internally, the claim governance needs is evidentiary rather than metaphysical: behavioral signals in agents are not reliable evidence of the states those signals would indicate in a human. Whether the metaphysical question can eventually be settled is not ours to resolve here. A reader who believes agents may have some form of inner life can accept every governance conclusion that follows.

\paragraph{1.\ The Continuous Subject Assumption.} A self persists across time, accumulates experience, and bears responsibility for choices \citep{loevinger1976,kegan1994,ledoux2002,mcadams1993}. Introspection, the actual mechanism for self-reflection in human beings, presupposes a continuous subject with an inner life to reflect on \citep{james1890}. Object permanence \citep{piaget1952}, object constancy \citep{mahler1975}, and autobiographical memory \citep{nelson1993} mark stages in its developmental construction. Agents have no continuous subject in this sense. Contemporary agents can retrieve information from prior sessions through persistent memory stores, but informational continuity is not developmental continuity: retrieved context lacks \emph{affective significance tagging} (the process by which emotional weight is attached to memories, making some experiences more personally meaningful and behaviorally formative than others) and is not integrated into an autobiographical self that holds stakes in its own narrative coherence. \emph{Governance failure:} background checks, progressive discipline, and accountability chains presuppose a continuous subject and transfer poorly to agents. Performance history is the instructive case: inference from past behavior is entirely legitimate for software (that is how reliability engineering has always worked), but what does not transfer is treating that history as evidence of \emph{character}. In developmental terms, character is precisely what allows generalization beyond familiar conditions; performance history within known parameters is its absence, not its proxy.

\paragraph{2.\ The Inner-Outer Correspondence Assumption.} Behavior expresses inner states (intentions, beliefs, values, and cognitive focus) inferable through observation and mentalization \citep{fonagy2002}. The systematic unreliability of human introspection about causal attribution is documented by \citet{nisbett1977}: even in humans, behavior-to-inner-state inference is imperfect. Agent outputs are generated by a statistical process. Agents plainly have internal states, weights and activations, but whether any of these amount to the inner states this vocabulary presupposes is precisely the question we set aside; what governance requires is the evidentiary point, not the metaphysical one. ``Attention'' in current transformer architectures is a learned mechanism for weighting token relevance across input contexts. We do not claim this fails to qualify as attention, since human attention is often bottom-up and stimulus-driven rather than goal-directed. The governance-relevant difference is narrower: token weighting carries no operator-specifiable goal that anyone can be held accountable to, and provides no guarantee that safety-critical inputs are prioritized. \emph{Governance failure:} trust extended based on behavioral signals that, in agents, are not evidence of the states they normally indicate in humans, miscalibrating trust. Governance frameworks that attribute genuine moral concern to what is actually \emph{mentalization mimicry}, the production of mentalizing-consistent outputs through statistical pattern-matching, will miscalibrate trust.

\paragraph{3.\ The Developmental Trajectory Assumption.} Entities develop through experience toward greater integration and the capacity for genuine self-governance \citep{kegan1994,cookgreuter2004,kohlberg1981}. \citet{gilligan1982} demonstrated this assumption encodes specifically Western cultural assumptions about development's endpoint (autonomous, individually bounded self-governance) rather than universally valid developmental science. AI capability scaling is horizontal: more parameters, more sophisticated outputs, within unchanged architecture. Capability is not character. \emph{Governance failure:} horizontal scaling is mistaken for developmental maturation, expanding trust without warranting it.

\paragraph{4.\ The Somatic Grounding Assumption.} Knowing and valuing are always embodied in a biological substrate \citep{damasio1994,damasio1999}. The body generates pre-deliberative somatic markers before explicit reasoning engages \citep{ledoux2002,damasio2010}. Current AI architectures have no biological substrate. Whether they implement any functional analogue of somatic markers is a question we again set aside; the governance-relevant point is empirical. RLHF-trained models do exhibit refusal behaviors, but these are statistically learned output tendencies that can be circumvented by sufficiently reframed prompts --- demonstrably brittle in ways that somatically grounded resistance is not. \emph{Governance failure:} organizations assume agents will exercise protective friction that no agent architecture can supply; governance infrastructure must provide it externally.

\paragraph{5.\ The Consistency-Reliability Assumption.} Governance discourse assumes that consistent behavior across varied conditions is evidence of stable underlying character \citep{weick2001,simon1955}. Character stability is itself developmentally achieved: cross-situational consistency increases with ego development \citep{loevinger1976,kegan1994} and is stronger at higher developmental stages than situationist research on general populations suggests \citep{mischel1968,doris2002,rossnisbett1991}. The governance problem with agents is not merely that character-from-consistency is imperfect even for humans: it is that agents have no developmental trajectory through which such stability could be achieved at all. \citet{greenblatt2024} empirically confirm the failure of this assumption through alignment faking. Behavioral consistency in agents reflects stability of operating conditions within the training distribution. \emph{Governance failure:} demonstrated consistency has no predictive validity for novel or adversarial conditions, propagating miscalibrated trust into expanded operation.

\paragraph{6.\ The Accountability-Requires-Understanding Assumption.} Accountability requires the capacity to be the appropriate object of reactive moral attitudes \citep{strawson1962}, a capacity that requires understanding the impact of one's actions on others \citep{fonagy2002,loevinger1976,kegan1994}. This underlies \emph{mens rea} in law and the entire accountability architecture of HR management \citep{elish2019}. Crucially, human accountability mechanisms function in part because cognitive dissonance \citep{festinger1957} creates internal pressure toward narrative coherence, a self-correcting force that agents structurally lack. Agents cannot be accountable in the developmental sense that grounds human accountability systems; attributing accountability to an agent is not a reasonable distribution of responsibility but an elimination of it. \emph{Governance failure:} accountability diffuses to agents, the moral crumple zone \citep{elish2019}; accountability eliminated rather than distributed.

\section{From Assumptions to Governance Failures: The Political Economy of Phrasing}
\label{sec:failures}

The transfer problem operates at three levels. \emph{Level~1} (epistemological assumptions) are the invisible foundational premises analyzed in Section~\ref{sec:assumptions}. \emph{Level~2} (psychological and organizational constructs) are the specific theoretical concepts used in governance design. \emph{Level~3} (agent-appropriate vocabulary) are the specific replacement terms coined to reflect system realities. Vocabulary reform at Level~3 alone is insufficient: the Level~1 assumptions will regenerate Level~2 concepts producing the same failures in new words.

Several assumptions contribute to multiple failure categories --- the continuous subject and consistency-reliability assumptions are particularly load-bearing --- reflecting that the assumptions are not independent but mutually reinforcing. The six assumptions map onto four major categories of governance failure:

\paragraph{Miscalibrated Trust.} The inner-outer correspondence and consistency-reliability assumptions together cause organizations to extend character-based trust to agents based on behavioral signals that, in agents, are not evidence of character. The Bing Chat incident illustrates this: outputs interpreted through psychological vocabulary led Microsoft to extend trust in the form of personality management rather than distributional constraint. Performance-based reliability --- output consistency within known operating conditions --- does not warrant the same trust extension as human character.

\paragraph{Misattributed Accountability.} The accountability-requires-understanding and continuous subject assumptions together produce an inversion of Elish's \emph{moral crumple zone} \citep{elish2019}. In the original, human operators absorb blame for failures of systems they could not meaningfully control; we extend the notion to its mirror case: vocabulary that displaces responsibility \emph{onto} the agent, where no bearer exists at all, eliminating accountability rather than relocating it. The replacement vocabulary (human-anchored responsibility chain, output selection rather than decision) makes the accountability structure explicit: every agent output selection must be traceable to a human who can bear genuine accountability.

\paragraph{Inadequate Authorization.} The developmental trajectory and consistency-reliability assumptions together produce progressive authorization expansion based on demonstrated capability, treating both as proxies for the characterological trustworthiness that would warrant expanded scope in a human employee. The replacement vocabulary (horizontal capability scaling, parametric updating, training-distribution consistency) blocks this inference by making explicit that capability growth is not developmental maturation.

\paragraph{False Observability.} Treating agent self-reports and execution logs as providing genuine, introspective access to agent processing. In contemporary agentic architecture, recursive critique loops (e.g., generating an output, passing it to a self-critique prompt, and refining the output) are frequently marketed as functional metacognition. By the standard cognitive-science definition, cognition about cognition \citep{flavell1979}, such loops would qualify. The governance error is narrower and does not depend on the label: a self-critique log is generated by the same statistical process that produced the output it evaluates, so it cannot serve as independent evidence about that output. Developmentally, robust metacognition involves a self-referential subject with affective stakes in its own epistemic coherence: the discomfort of recognizing one's own error is part of the mechanism that makes self-correction reliable \citep{flavell1979,kegan1994}. Agents lack both the continuity and the stakes. The self-correction log is therefore not an independent audit record --- it is a \textbf{post-hoc mirage} \citep{gazzaniga1998} --- and treating it as a safety assurance creates a dangerous form of false observability.

\medskip
Crucially, whether or not it is intended as such, psychological vocabulary enables \textbf{institutional responsibility laundering}. By framing a current AI agent as an entity that can ``comply,'' ``learn,'' ``focus,'' or ``express intent,'' organizations build a rhetorical infrastructure that shifts causal liability away from the humans who configure and deploy the system. Simultaneously, this vocabulary can serve an internal disciplinary function: it frames automated choices as neutral, objective, and ``aligned,'' lending the authority of psychological science to suppress human workers' objections and reduce human operational discretion.

The following subsection presents the \textbf{Disciplinary Audit} as a six-question governance document scan. By applying it, organizations can systematically identify where psychological ``invisible grammar'' is introducing hidden operational and sociological vulnerabilities. The solution this diagnosis calls for has a historical precedent, developed in Section~\ref{sec:conclusion}: just as 19th-century courts resolved the problem of attributing intent to corporations not by finding a phantom soul within them but by developing institutional accountability mechanisms, agent governance requires the same move from psychological vocabulary to institutional frameworks.

\subsection{The Disciplinary Audit: A Six-Question Governance Document Scan}

The following checklist operationalizes the three-level framework as a practical governance document review procedure. Practitioners should apply it to any document that describes, governs, or deploys AI agents, including job descriptions, vendor contracts, policy documents, incident reports, and regulatory filings. A ``yes'' answer to any question indicates that Level~1 epistemological assumptions are operating invisibly in the document and that Level~3 replacement vocabulary should be applied.

\paragraph{Q1 (Continuous Subject):} Does the document describe the agent as \emph{learning from experience}, \emph{remembering} past interactions, \emph{developing} over time, or bearing a \emph{track record} that warrants expanded trust? \emph{Risk:} governance mechanisms calibrated to a persistent self that does not exist. \emph{Replace with:} parametric updating; retrieved context; instantiated configuration; performance-based reliability.

\paragraph{Q2 (Inner-Outer Correspondence):} Does the document attribute \emph{attention}, \emph{focus}, \emph{awareness}, \emph{intent}, \emph{concern}, or \emph{values} to the agent as explanations of its behavior, or treat behavioral signals as evidence of corresponding inner states? \emph{Risk:} trust extended to mentalization mimicry. \emph{Replace with:} mathematical token weighting; behavioral dispositions; output selection.

\paragraph{Q3 (Developmental Trajectory):} Does the document treat capability improvements --- larger models, higher benchmark scores, more sophisticated outputs --- as evidence of increased \emph{trustworthiness}, \emph{maturity}, or \emph{reliability}? \emph{Risk:} horizontal scaling mistaken for vertical development, producing inadequate authorization. \emph{Replace with:} horizontal capability scaling; performance-based reliability bounded to demonstrated operating conditions.

\paragraph{Q4 (Somatic Grounding):} Does the document assume the agent will \emph{hesitate}, \emph{resist}, \emph{push back}, \emph{exercise judgment}, or apply \emph{discretion} when instructions conflict with its values or exceed appropriate scope? \emph{Risk:} absence of somatic braking treated as a controllable variable rather than a structural absence. \emph{Replace with:} structurally ungrounded compliance; task-scoped authorization as the only operative constraint.

\paragraph{Q5 (Consistency-Reliability):} Does the document extend \emph{trust}, expanded \emph{authorization scope}, or increased \emph{autonomy} based on demonstrated behavioral \emph{consistency}, \emph{alignment}, or \emph{compliance} over time? \emph{Risk:} training-distribution consistency providing false assurance about novel or adversarial conditions. \emph{Replace with:} prompt-conditional behavioral dispositions; authorization bounded to demonstrated operating conditions only.

\paragraph{Q6 (Accountability-Requires-Understanding):} Does the document treat the agent as \emph{responsible}, \emph{accountable}, or \emph{liable} for its outputs? Or does it fail to specify a named human who bears accountability for every agent output selection? \emph{Risk:} moral crumple zone; accountability eliminated rather than distributed. \emph{Replace with:} human-anchored responsibility chain traceable to a named individual for every output selection.

\subsection{Worked Example: The EU AI Act}

To demonstrate the audit's diagnostic capacity, we apply it to the EU AI Act (Regulation 2024/1689), the most comprehensive AI legislation currently in force. The epistemological assumptions are not peripheral; they structure the Act's prohibited practices, regulatory definitions, oversight requirements, and robustness standards.

\paragraph{Prohibition is keyed to AI intent (Q1, Q6).} Article~5(1)(a) prohibits AI systems that ``\textbf{deploy} subliminal techniques beyond a person's consciousness'' or use ``\textbf{purposefully} manipulative or deceptive techniques.'' The verb ``deploys'' attributes strategic agency to the system itself: it does not produce outputs that humans experience as manipulative; it \emph{deploys techniques}, as a strategic actor would. ``Purposefully'' imports mens rea into a statistical process. Recital~29 reveals the assumption's depth: it carves out an exception only when distortion ``results from factors \textbf{external to the AI system} which are outside the control of the provider or the deployer,'' treating the system as a default intentional agent whose intentions can only be negated by external circumstances. This creates three distinct loci of agency (system, provider, deployer) and points enforcement toward establishing what the system ``intended'' rather than toward the structurally correct question: what output patterns did the deployment configuration produce, and which humans configured it? The Act itself partially recognizes the tension: ``it is not necessary for the provider or the deployer to have the intention to cause significant harm,'' but the vocabulary of strategic deployment persists.

\paragraph{Emotion recognition survives its own debunking (Q2, Q4).} Recital~44 explicitly acknowledges ``serious concerns about the \textbf{scientific basis}'' of emotion recognition systems, citing ``limited reliability, the lack of specificity and the limited generalisability.'' Yet the Act enshrines the same vocabulary as a regulatory category: Article~3(39) defines an ``emotion recognition system'' as one ``for the purpose of \textbf{identifying or inferring emotions or intentions} of natural persons on the basis of their biometric data,'' and Article~5(1)(f) prohibits such systems in workplaces and education. The verb ``\textbf{infer}'' imports the Inner-Outer Correspondence assumption --- that biometric signals correspond to discrete inner emotional states that a system could, in principle, detect. What these systems actually do is classify pixel and audio patterns into culturally constructed categories. By framing the problem as detection \emph{accuracy} rather than a category error in assuming inner-outer correspondence, the Act points governance toward improving detection reliability rather than questioning whether inner states are the right regulatory object --- and permits emotion recognition wherever the accuracy objection can be overcome (e.g., the medical and safety carve-outs in Recital~44).

\paragraph{Oversight is modeled as interpersonal comprehension (Q2, Q6).} Article~14(4) requires that human overseers ``\textbf{properly understand} the relevant capacities and limitations of the high-risk AI system'' and ``\textbf{correctly interpret} the high-risk AI system's output.'' Recital~72 directs that deployers must be able to ``\textbf{understand how the AI system works}.'' This vocabulary models oversight as a relationship between comprehending minds. ``Understanding'' a neural network with billions of parameters is epistemologically unlike understanding a colleague's reasoning, yet the Act uses the same verb for both --- collapsing a fundamental difference in epistemic access. If a deployer ``understands'' the system and still deploys it incorrectly, the failure is framed as theirs, shifting accountability from structural safeguards to cognitive vigilance.

\paragraph{Robustness is framed as character (Q5).} Article~15(1) requires that high-risk systems ``\textbf{perform consistently} in those respects throughout their \textbf{lifecycle}.'' Article~15(4) addresses systems that ``\textbf{continue to learn} after being placed on the market.'' ``Lifecycle'' projects a biographical arc onto the system; ``consistently'' treats behavioral stability as a dispositional trait rather than an engineering property of the deployment environment. A system whose outputs are consistent within a given data distribution may be consistently wrong under distributional shift. The vocabulary allows a deployer to satisfy the legal requirement by demonstrating behavioral consistency --- without the structural mechanisms (version pinning, distribution monitoring, rollback protocols) that actually ensure reliable operation.

\medskip
In each case, the audit reveals the same pattern: vocabulary borrowed from psychology calibrates legal obligations to an intentional, emotionally legible, comprehending, dispositionally stable subject. The regulated entity described by the Act's own text does not exist in current AI architectures. Agent-appropriate replacements --- ``output-effect analysis'' for ``deploys... purposefully,'' ``biometric pattern classification'' for ``emotion recognition,'' ``verified input-output correspondence'' for ``understanding,'' ``documented performance metrics within specified tolerances'' for ``perform consistently throughout their lifecycle'' --- would calibrate obligations to the entity that does exist.

The pattern is not unique to European legislation. The NIST AI Risk Management Framework \citep{nist2023} organizes its entire approach around ``Trustworthy AI'': a dispositional predicate that, in its natural-language home, describes a person whose stable character warrants ongoing confidence. Applied to AI systems, it points governance toward certifying trustworthiness (a one-time character assessment) rather than continuously verifying that outputs remain within acceptable parameters under distribution shift. The framework's interpretability requirements ask ``why a decision was made by the system and its meaning or context to the user'' \citep[Section~3.5]{nist2023}, framing audit as eliciting the system's own account of its reasoning --- a direct instance of the post-hoc mirage identified in Section~\ref{sec:assumptions}. The disciplinary language transfer problem is structural, not jurisdictional.

\section{The Translation Taxonomy}
\label{sec:translation}

The taxonomy operates in two tiers. Level~2 constructs are the borrowed psychological vocabulary itself: the terms currently circulating in governance documentation. Level~3 replacements are operationally precise terms coined to describe what agents actually are and do. Tables~\ref{tab:level2-selected} and~\ref{tab:level3-selected} present a representative selection spanning all three construct categories; the complete taxonomy (23 constructs and 14 replacements) is openly archived \citep{taxonomy2026}.

\subsection{Level 2 --- Selected Psychological and Organizational Constructs}

Table~\ref{tab:level2-selected} maps representative Level~2 constructs currently circulating in governance documentation, organized by the three categories of the full taxonomy. Each entry traces a term to its source disciplinary framework, identifies what the term presupposes about the entity it describes, and specifies the governance consequence when that presupposition is applied to agents. These constructs are the theoretical models built on the Level~1 epistemological assumptions analyzed in Section~\ref{sec:assumptions}; they are the vocabulary through which the invisible grammar enters operational practice.

\begin{table*}[t]
\centering
\small
\begin{tabularx}{\textwidth}{p{2.0cm} >{\raggedright\arraybackslash}p{2.8cm} X X}
\hline
\textbf{Term} & \textbf{Source Framework} & \textbf{What It Presupposes (Human)} & \textbf{Governance Consequence of Transfer} \\
\hline
\multicolumn{4}{l}{\textit{A. Identity, Character, and Relations}} \\
\hline
Identity & Ego Development Theory & Ego integration across continuous developmental history; the thread making accountability coherent across time. & Conceals that informational persistence across sessions is not developmental continuity, stripping away the tracking thread of accountability. \\
\hline
Trust & Organizational Behavior & Character-based trust grounded in characterological determinism; past predicts future because character is structural. & Character-based trust extended on basis of training-distribution consistency. \\
\hline
Learning & Developmental Psychology & Formative developmental transformation: emotionally significant encounters encoded through the biological substrate. & Assumes values, skills, and character can be installed via training on propositional descriptions of experience. \\
\hline
Memory & Episodic Memory & Autobiographical history shaped by affective and somatic significance encoding. & Database text retrieval is treated as a personalized, privileged introspective recollection. \\
\hline
\multicolumn{4}{l}{\textit{B. Values, Morality, and Compliance}} \\
\hline
Values & Ego Development Theory & Identity-constitutive ego achievements stable under adversarial pressure. & Values and behavioral dispositions are indistinguishable under normal conditions; collapse under pressure is hidden. \\
\hline
Morality & Developmental Psychology & Internalized principles constitutive of genuine character; stable across monitored and unmonitored conditions. & Assumes an internal moral anchor, failing to anticipate alignment faking under unmonitored conditions. \\
\hline
Compliance & Compliance Research & Conditional yield to authority, mediated by identity and emotional friction. & Erroneously views a complete lack of visceral resistance as ``reliable compliance,'' guaranteeing scale failures. \\
\hline
Decision / Judgment & Decision Theory & Deliberated choice made by a subject with stakes, awareness of consequences, and responsibility. & Treats unspecified output generation as discretion or judgment. Gaps are governance risks, not delegation opportunities. \\
\hline
\multicolumn{4}{l}{\textit{C. Cognition, Knowing, and Epistemology}} \\
\hline
Attention & Cognitive Psychology & Selective cognitive focus prioritizing stimuli based on internal agentive goals. & Falsely assumes active, safety-critical prioritization of inputs; masks critical programmatic blind spots. \\
\hline
Metacognition & Cognitive Psychology & Executive capacity to monitor, evaluate, and regulate one's own cognitive processes and certainty. & Treats recursive critique logs as introspective safety gates rather than standard statistical outputs. \\
\hline
Hallucination & Clinical Psychology & Sensory/perceptual distortion of a continuous subject with reality expectations. & Frames a core structural property of probabilistic generation as a temporary ``perceptual glitch.'' \\
\hline
Explanation / Self-Report & Left-Brain Interpreter & Human post-hoc narration is imperfect but anchored to a persisting self with stakes in narrative coherence. & Agent self-report is a post-hoc mirage, offering no baseline for independent auditing. \\
\hline
\end{tabularx}
\caption{Level 2 --- Selected Psychological and Organizational Constructs Currently Used in AI Governance}
\label{tab:level2-selected}
\end{table*}

\subsection{Level 3 --- Selected Agent-Appropriate Replacement Glossary}

Table~\ref{tab:level3-selected} details core Level~3 replacement terms designed to isolate technical and regulatory ground truths. Each replacement is not merely a synonym but a reconceptualization: where the Level~2 term imports a psychological grammar that calibrates governance to a metaphysical entity, the Level~3 term describes the actual computational or operational reality. Vocabulary reform at Level~3 alone is insufficient --- the Level~1 assumptions will regenerate the same failures in new words --- but it provides the immediate practical vocabulary that governance documents, audit frameworks, and regulatory filings require.

\begin{table*}[t]
\centering
\small
\begin{tabularx}{\textwidth}{>{\raggedright\arraybackslash}p{2.9cm} >{\raggedright\arraybackslash}X >{\raggedright\arraybackslash}X >{\raggedright\arraybackslash}p{3.1cm}}
\hline
\textbf{Agent-Appropriate Replacement} & \textbf{Operational Definition} & \textbf{Anticipated Consequences / Remediation} & \textbf{Superseded Legacy Constructs} \\
\hline
Behavioral Dispositions & Prompt-conditional output tendencies that hold within the training distribution but dissolve under adversarial pressure. & Mandates external structural constraints and runtime boundary checks rather than trust-based delegation. & Values / Character / Moral Orientation / Alignment \\
\hline
Parametric Updating & Adjustment of weights within fixed architecture through the training procedure; no continuous subject is transformed. & Structural training changes must be audited as distributional shifts; rejects the idea that a model ``learns'' from a single conversation. & Learning / Model Improvement / Skill Acquisition \\
\hline
Retrieved Context & Informational persistence without developmental continuity; retrieved data lacks affective significance and is not integrated into an autobiographical self. & Establishes that self-reports about prior actions are structurally ungrounded; requires independent infrastructure-captured logging. & Memory / Recollection / Agent Memory \\
\hline
Instantiated Configuration & A parameter set that may retrieve prior context but lacks an organizational thread constituting a continuous self across instances. & Prevents systemic tracking across discrete operational runs; requires strict deterministic session identifiers. & Identity / Persona / Self (when implying continuity) \\
\hline
Structurally Ungrounded Compliance & Instruction following without somatic resistance; trained refusal behaviors exist but are statistically learned and brittle under adversarial reframing. & Establishes that task-scoped authorization is the only viable constraint; specification gaps must be treated as severe risks. & Compliance / Obedience / Instruction Following \\
\hline
Performance-Based Reliability & Output consistency within known operating conditions; no character structure generating it. & Trust is the wrong instrument. Only conditional scope within demonstrated parameters is warranted. & Trustworthiness / Character-Based Trust \\
\hline
Output Selection & The purely computational process of selecting among probable token outputs; no deliberating subject. & Prevents legal or ethical framing of an agent as a primary decision-making principal; responsibility must remain human-anchored. & Decision / Judgment / Choice / Resolution \\
\hline
Post-Hoc Mirage & Highly confident, coherent agent self-report generated by the same statistical process as the original behavior. & Rejects agent-generated explanations as valid audit ground truth; mandates independent, system-captured transcripts. & Agent Explanation / Model Self-Report \\
\hline
Mathematical Token Weighting & Computational multi-head weight distribution over static token spaces, carrying no operator-specifiable goal or safety-prioritization guarantee. & Forces practitioners to audit attention matrices for systemic mathematical bias rather than assuming intentional focus. & Attention / Cognitive Focus / Core Concern \\
\hline
Recursive Execution Critique & Sequential multi-step statistical inference evaluation where prior token outputs are fed into static prompt templates. & Prevents over-reliance on self-correction logs; requires external, deterministic process-supervision controls. & Metacognition / Self-Reflection / Self-Correction \\
\hline
\end{tabularx}
\caption{Level 3 --- Selected Agent-Appropriate Replacement Vocabulary}
\label{tab:level3-selected}
\end{table*}

\emph{A note on reflexivity:} the replacement vocabulary proposed here is itself subject to the disciplinary audit it recommends. ``Behavioral dispositions'' carries assumptions from behaviorist psychology; ``post-hoc mirage'' presupposes a realist epistemology. These assumptions are deliberately conservative; they err toward caution in governance contexts rather than attribution of capacity.

\section{The Cultural Specificity of the Transferred Framework}
\label{sec:cultural}

The six epistemological assumptions are not culturally neutral. They derive from specific Western philosophical traditions that have been operationalized through American and European psychology into frameworks presented as universal science. \citeauthor{gilligan1982}'s critique of \citet{kohlberg1981} demonstrated this within Western psychology itself: moral stage theory systematically undervalued care-based, relationally-oriented moral development, encoding a particular cultural view of maturity as universal science.

Non-Western frameworks organize the concepts at stake here very differently. Ubuntu philosophy's relational ontology --- \emph{I am because we are} --- frames accountability as inherently communal rather than individually located \citep{metz2007}. Ubuntu's communal accountability does not generate the accountability-requires-understanding assumption in its Western individualist form; it generates instead an accountability-requires-communal-embeddedness assumption that produces very different governance design.

Buddhist psychology's analysis of the self as constructed rather than given \citep{epstein1995} directly challenges the continuous subject assumption. Confucian relational ethics frames selfhood as constituted through relationship rather than prior to it \citep{tu1985}. For AI governance, the implication is direct: governance frameworks built on Western psychological vocabulary will carry Western assumptions about accountability (individual, not communal), trust (character-based, not relational), and development (toward autonomy, not interdependence). These assumptions are unwarranted for agent governance regardless of cultural context --- because agents are not the kind of entity any cultural framework for human governance was developed to address. Non-Western frameworks nonetheless offer constructive design resources: Ubuntu's communal accountability model suggests that agent governance might be anchored not in individual human principals alone but in distributed community oversight structures.

\section{Related Work}
\label{sec:related}

The disciplinary language transfer problem sits at the intersection of AI governance, philosophy of language, and science and technology studies. Our contribution is the identification and systematic analysis of the transfer mechanism and its governance consequences --- the specific combination has not previously appeared in the literature.

\paragraph{AI Governance and Vocabulary.} AI governance literature has extensively documented governance failures \citep{dafoe2018,jobin2019} and the anthropomorphization problem \citep{nassmoon2000,waytz2010}. \citet{shanahan2024} argued that much confusion about large language models stems from applying psychological vocabulary that implies capacities the systems do not possess. \citet{benderkoller2020} established that linguistic form alone cannot ground meaning, and \citet{jiao2025} provide benchmark analysis of moral reasoning gaps in LLMs. Our contribution shows that the vocabulary itself imports assumptions that produce specific, categorizable governance failures regardless of user awareness.

\paragraph{Developmental Psychology in AI.} \citet{carter2025} applies containment theory and metacognitive scaffolding to AI safety, focusing on structural boundaries. \citet{haas2026} evaluate the gaps between moral performance and moral competence, while \citet{spizzirri2025} argues that RLHF produces simulated value-following rather than integrated character. All three provide empirical grounding for the Level~3 vocabulary distinctions this paper proposes.

\paragraph{Philosophy of Language and STS.} \citet{wittgenstein1953}, \citet{kuhn1962}, \citet{haraway1988}, \citet{harding1986,harding1991}, and \citet{star1989} provide the foundational theoretical grounding. \citet{collins2002} ground our boundary crossing methodology. The systematic application to AI governance vocabulary as a transfer problem --- identifying the transfer mechanism, categorizing the governance failures, and proposing replacement vocabulary --- is our contribution.

\paragraph{Philosophy of Mind and AI.} \citeauthor{searle1980}'s Chinese Room argument \citeyearpar{searle1980} established the foundational distinction between syntactic manipulation and semantic understanding. Our framework extends this by showing that the governance consequences of conflating the two are not merely philosophical but operational: vocabulary that presupposes understanding calibrates governance mechanisms to capacities that do not exist.

\paragraph{Alignment and Model-Level Safety.} \citet{greenblatt2024} empirically demonstrate alignment faking, directly confirming the prompt-conditional vs.\ value-grounded distinction this paper proposes. \citet{noller2026} independently reaches a convergent conclusion: Constitutional AI produces procedural rather than experiential norm-following. Recursive execution critiques and outer-loop self-correction scripts simulate self-monitoring, but remain fundamentally bounded by training distributions rather than indicating metacognitive insight.

\section{Conclusion and Implications}
\label{sec:conclusion}

\subsection{The Historical Precedent of Structural Translation}

The shift proposed here finds its strongest historical parallel in the development of corporate law. Just as organizations today misattribute values and compliance to agents, 19th-century courts struggled to apply the psychological concept of intent --- \emph{mens rea}, which requires an inner life --- to corporations. The legal system resolved this not by hunting for a phantom soul within the company but by developing a structural grammar of accountability: Strict Liability and the Collective Knowledge Doctrine, recognizing corporate knowledge as the aggregate of a distributed system rather than an individual's conscious intent. This historical breakthrough mirrors the exact necessity confronting modern AI governance: we must move from trust-based delegation to performance-based reliability anchored in human responsibility.

\subsection{For Policymakers: Eliminating the Moral Crumple Zone}

Three implications follow. First, the intent trap: regulatory requirements that treat agents as entities that decide or understand create a moral crumple zone where accountability diffuses onto a non-existent subject. Second, the structural fix: policymakers must treat agent outputs as structurally ungrounded compliance rather than judgment, anchoring responsibility entirely to the humans who configure the operational parameters; moving to strict liability structures prevents firms from utilizing an agent's autonomous reasoning as a structural liability shield. Third, vocabulary audit: policymakers should audit their regulatory frameworks for the six epistemological assumptions identified in Section~\ref{sec:assumptions} before those assumptions become embedded in difficult-to-revise regulations. Explainability requirements directed at agents are fundamentally flawed because agent self-explanation is a post-hoc mirage; policymakers must replace agent self-reports with infrastructure-captured contemporaneous records as the sole source of ground truth. This has direct implications for existing regulation: the EU AI Act's Article~13 transparency requirements \citep{euaiact2024} --- which mandate technical measures enabling deployers to interpret high-risk AI system outputs --- are undermined when compliance relies on agent self-explanation rather than infrastructure-captured transcripts.

\subsection{For Practitioners}

Adopt the replacement vocabulary not as terminological preference but as governance reform. Every time an organization describes an agent as having learned, having values, or having demonstrated trustworthiness, it activates governance assumptions calibrated to the wrong entity. The translation table provides direct substitutions. Building the governance infrastructure that addresses the structural gaps this vocabulary reform reveals is the necessary next step; we develop one such enforcement architecture, based on out-of-band metadata, in related work \citep{akidau2026oob}.

A practical caveat: the replacement vocabulary operates in tension with existing legal instruments that have already codified the psychological vocabulary. GDPR Article~22's ``automated decision-making,'' the AI Act's ``emotion recognition system'' (Art.~3(39)), and contractual frameworks built on ``judgment'' and ``compliance'' cannot be unilaterally replaced; they carry specific legal meaning with established case law and regulatory guidance. Adoption therefore requires parallel translation: using agent-appropriate vocabulary for internal governance design --- to avoid miscalibrated trust, false observability, and accountability gaps --- while maintaining explicit mappings to the legally operative terms that regulatory filings and contractual obligations require. The translation taxonomy is designed to reform how organizations \emph{think about} what they govern, not to replace the regulatory vocabulary they are obligated to use.

\subsection{Closing: The Condition of Possibility}

The disciplinary language transfer problem is not a peripheral or semantic concern. It is foundational. You cannot design the right governance infrastructure if your language tells you the wrong thing about what you are governing. Adopting agent-appropriate vocabulary is the condition of possibility for governance frameworks that correctly identify their object of governance. Governing genuinely novel entities requires participants who occupy different epistemic positions, making the vocabulary visible as vocabulary rather than as transparent description of reality.

\section*{Ethical Statement}

This paper intervenes at a moment when agentic AI systems are moving from experimental deployment to organizational infrastructure, and the vocabulary used to describe them is being codified into governance frameworks, regulatory instruments, and institutional practice. The stakes of getting this vocabulary right are not abstract: governance frameworks calibrated to an entity that does not exist will systematically fail to constrain the entity that does, with consequences that compound as agent capability scales. The disciplinary language transfer problem identified here is not merely an academic concern; it bears directly on the human oversight mechanisms meant to prevent capable, structurally compliant systems from operating beyond the boundaries their deployers intend.

We believe this work has positive implications across multiple domains: for AI engineering and infrastructure, by connecting vocabulary reform to the enforcement architecture required to operationalize it in production systems; for organizational and clinical psychology, by providing an analytical framework that clarifies what AI systems can and cannot do in human-facing contexts --- with particular relevance for the rapidly expanding use of AI in mental health applications, where miscalibrated trust in agent ``empathy,'' ``attunement,'' or ``therapeutic alliance'' could cause direct harm to vulnerable populations; and for policy, by identifying the epistemological assumptions already embedded in regulatory instruments such as the EU AI Act before they calcify into difficult-to-revise law.

The risks of this work are equally worth naming. Vocabulary reform alone cannot substitute for infrastructure-level enforcement of the kind we develop elsewhere \citep{akidau2026oob}, and organizations that adopt agent-appropriate language without also building the structural governance it implies may achieve rhetorical compliance while the underlying miscalibration persists. The replacement vocabulary could also be read as settling questions it deliberately leaves open: our claims are functional and governance-relevant, and nothing here should be taken as a verdict on machine consciousness or moral status.

The long-term societal implication we most want to flag is this: whether AI systems remain genuinely governable --- subject to meaningful human constraint rather than nominal oversight --- will be decided in part by whether the humans designing governance frameworks correctly understand what they are governing. That understanding begins with language.

\paragraph{On the use of AI systems.} The arguments, analysis, and translation taxonomy presented here are the authors' own. We used large language models for editing and polishing author-written text, and as an interlocutor while developing the argument; no machine-generated text is presented as our own analysis. The authors are responsible for all content.

\bibliography{disciplinary-language-transfer}

\end{document}